# Status of Telescope Fabra ROA at Montsec: Optical Observations for Space Surveillance & Tracking

**Fors O.**
*Observatori Fabra, Reial Acadèmia de Ciències i Arts de Barcelona*
*Dept. d'Astronomia i Meteorologia and Institut de Ciències del Cosmos (ICC), Universitat de Barcelona (UB/IEEC)*

**Montojo F.J.**
*Real Instituto y Observatorio de la Armada (ROA)*

**Núñez J.**
*Observatori Fabra, Reial Acadèmia de Ciències i Arts de Barcelona*
*Dept. d'Astronomia i Meteorologia and Institut de Ciències del Cosmos (ICC), Universitat de Barcelona (UB/IEEC)*

**Muiños J.L.**
*Real Instituto y Observatorio de la Armada (ROA)*

**Boloix, J.**
*Real Instituto y Observatorio de la Armada (ROA)*

**Baena R.**
*Observatori Fabra, Reial Acadèmia de Ciències i Arts de Barcelona*
*Dept. d'Astronomia i Meteorologia and Institut de Ciències del Cosmos (ICC), Universitat de Barcelona (UB/IEEC)*

**Morcillo R.**
*Real Instituto y Observatorio de la Armada (ROA)*

**Merino M.**
*Observatori Fabra, Reial Acadèmia de Ciències i Arts de Barcelona*
*Dept. d'Astronomia i Meteorologia and Institut de Ciències del Cosmos (ICC), Universitat de Barcelona (UB/IEEC)*

1. ABSTRACT

The telescope Fabra ROA at Montsec (TFRM) is a 0.5m f/1 refurbished Baker-Nunn Camera (BNC) operated by a collaboration between the Fabra Observatory - Royal Academy of Arts and Sciences of Barcelona and the Spanish Navy Observatory (ROA), and installed at Montsec Astronomical Observatory (Spain).

Among other capabilities, its CCD FoV (4.4degx4.4deg), the telescope tracking at arbitrary RA and DEC rates, and the CCD shutter commanding at will during the exposure are specially remarkable for Space Surveillance and Tracking (SST) observational programs.

On Feb 2011, the TFRM participated, in the CO-VI third run satellite tracking campaign of the ESA SST/Space Surveillance Awareness Preparatory Program (SST/SSA-PP). During this multi-asset 7-day campaign the TFRM conducted systematic observations of artificial satellites which yielded to the determination of 1137 accurate position measurements. Since Feb 2011, the TFRM is observing in remote and fully unattended robotic modes under commissioning status.

A summary of the results of the ESA CO-VI SST optical observational campaign and insights of other SST-like observations in process will be presented.

## 2. TFRM ORIGINAL AND CURRENT CAPABILITIES

One of the 21 BNCs constructed by the Smithsonian Institution during the early Space Age was installed in 1958 at the Spanish Navy Observatory (ROA) in San Fernando (Spain). These were designed and constructed for optical satellite tracking purposes, so the highest optical and mechanical requirements were met at those days by Perkin & Elmer and Boller & Chivens.
In short, the BNC was an f/1 0.5m three-element lenses corrector cell (modified Schmidt telescope) with a 0.78m diameter primary mirror. A 20μm spot size was guaranteed throughout a 30ºx5º FoV over a photographic film. Once new technologies enabled new satellite tracking facilities as GEODSS, the BNC program was cancelled and the one in San Fernando was donated to ROA, where it has been maintained in excellent state of conservation.

Over the years, with the maturity of information and communication technologies, data analysis, telescope networks, CCD devices, domotics, and observatory control software, robotic astronomy has reached a level of development never seen before. Very complex observation requests can be scheduled in advance with more and better scientific outcome and the same reliability as the classical in situ observer mode.
The habitual profile of a robotic telescope is a brand new one which has been manufactured according to the required specifications. This usually brings us up to 2.5m in diameter, ranges f/3 to f/10 in focal ratio, commonly a Ritchey-Chrétien design, and with light and fast mounts. However, there is no precedence of a fully robotic telescope with f/1 focal ratio and moderate aperture as BNC's. The aim to fill that niche in favor of SST observations, the BNC excellent optical and mechanical specifications, and the previous successful experiences of adapting other BNCs to CCD imaging (Phoenix, APT, RAO BNC), were the main drivers to undertake the refurbishment project of the BNC at ROA.
References [1] and [2] provide extensive descriptions of the origins and the refurbishment project of the BNC, now installed at the Montsec Astronomical Observatory (OAdM), and renamed as Telescope Fabra ROA at Montsec (TFRM).

After 4 years and a modest budget of 500kEUR over the whole period, the refurbishment project was successfully completed, and the TFRM was inaugurated on Sep 16 2010 as seen in Fig.1.

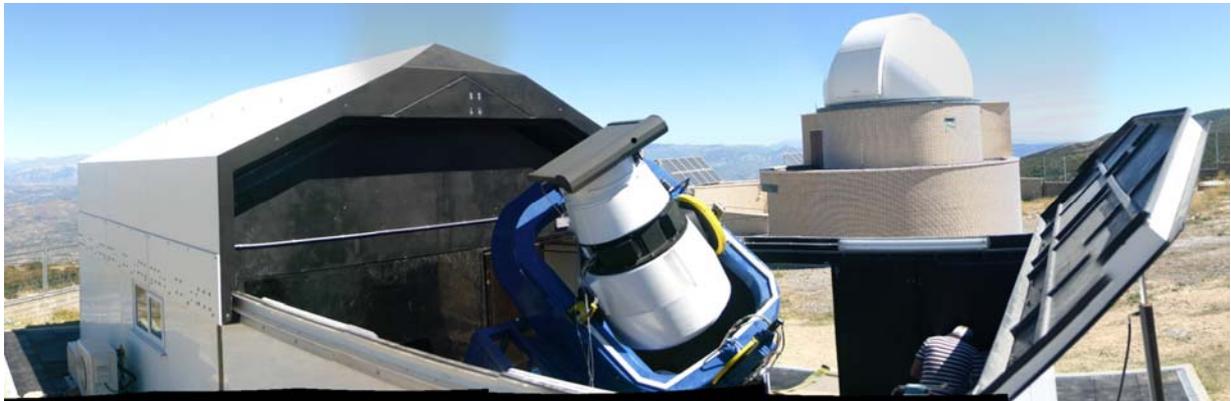

Fig. 1. Telescope Fabra ROA at Montsec (TFRM) at its current observing site, the Astronomical Observatory of Montsec (Catalonia, Spain). See reinforced glass-fiber enclosure with sliding roof open and South gable wall opening.

Just to enumerate the most remarkable specifications of the TFRM: f/0.96 focal ratio; a flat FoV of 6.25º in diameter achieved thanks to the addition of two lenses (meniscus and field flatenner) as corrective optics; a 4Kx4K Finger Lakes Instrumentation PL16803 CCD camera of 9μm-size (3.88arcsec) pixels covering a FoV of 4.4ºx4.4º with a 20μm spot size which encircles at least 80% of energy throughout the FoV; new spider vanes and CCD focus system is practically athermal with a focus accuracy of ±10μm; CCD camera cooled by a glycol recirculation chiller; servo driven equatorial mount with a maximum slewing speed of 2º/s, and CCD shutter operated at will during exposure with 0.1ms accurate GPS timestamp.
To properly orchestrate all above and other observatory devices in both remote and robotic modes, the TFRM is commanded by INDI, an state-of-the-art observatory control software created by Elwood C. Downey [3] and specifically customized for the TFRM with a suite of Java remote clients and robotic schedulers. INDI target tracking tools offers a functionality which is of particular utility for SST observations: the control software continuously computes the object coordinates by just providing its TLE elements, as seen in Fig.2.

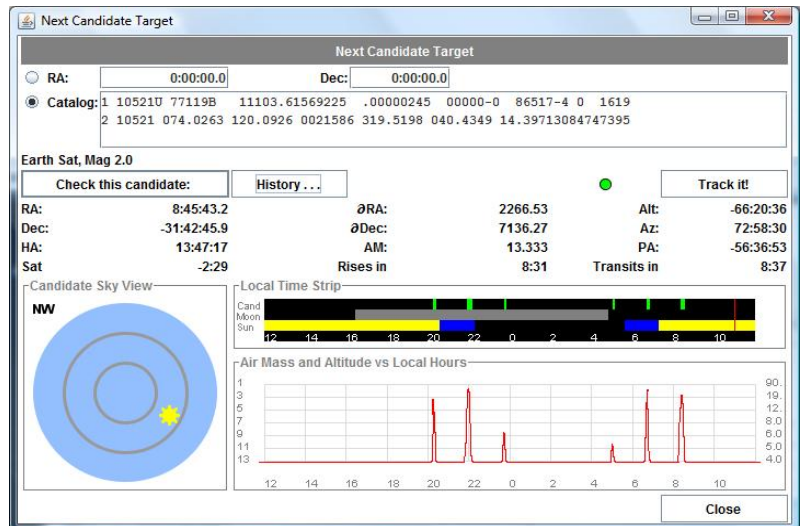

Fig.2. Target selection window. A celestial position can be input by coordinates, catalogue name or by its orbital elements (TLEs or .edb format). The picture shows a satellite tracking sequence by TLE selection. Note the visibility Time strip from the station and the logaritmical Air Mass vs. Local Time graph.

## 3. STATUS AND OBSERVATIONS UNDER COMMISSIONING PERIOD

INDI remote and robotic control software was tuned by the end of Dec 2010. Since then, the TFRM has been intensively tested under commissioning period. The observatory has been continuously operative and conducted numerous remote and robotic observations in the framework of different programs, such as exoplanets detection by transit technique, SST, NEOs, and X-ray and γ–ray optical counterparts identification and monitoring.

The following combination of TFRM instrumental specifications make an efficient facility to track and survey from faint GEOs to fast LEOs objects: moderately deep limiting magnitude (V~20mag) with 30s integration time, huge FoV (4.4º x 4.4º) free of optical aberrations, capability of tracking simultaneously in RA and DEC at arbitrary rates and, of triggering several times the CCD shutter at will during an exposure. Astrometric accuracies of 0.25arcsec for point-like objects and of 0.50arcsec for star trails were typically obtained (the latter ones by using SExtractor).

In collaboration to the International Scientific Optical Network (ISON) [5] a number of targets in the GEO ring area have been observed in order to assess the quality of such trailed images with APEX-II software [6]. Preliminary results are encouraging, both in terms of the astrometric accuracy improvement compared to the one obtained with SExtractor, and the automatic detection of faint GEO objects.

## 4. FIRST SST OBSERVATIONS

In this Section, we show how the TFRM refurbishment project has enabled this telescope to efficiently conduct both space surveillance and tracking programs, for Geostationary (GEO), Medium (MEO) and Low Earth Orbit (LEO) objects.

As an example of TFRM in GEOs tracking performance, this telescope participated, as an informal partner, in the third run of the CO-VI satellite tracking campaign, in the framework of the ESA SST/Space Surveillance Awareness Preparatory Program (SST/SSA-PP). In this program telescopes and radars from different countries (UK, Sweden, Switzerland, Cyprus and Spain) also participated. In the case of Spain, up to three observing stations (counting the TFRM) were involved. The satellite positions of every asset were submitted to the coordinating office at

Astronomical Institute University of Bern (AIUB), which reported the global results of the campaign [4]. As a result of that participation, the TFRM has already been integrated in the ESA SST program.

From 30 Jan 2011 to 7 Feb 2011, staff from the Fabra Observatory and the ROA conducted systematic observations of artificial satellites to determine 1137 accurate position measurements. This program is designed to gather information about the current European observational assets, to guarantee the security of spatial navigation at different orbits, and to lengthen their lifetime in space. Although under in-situ human supervision, this was the first official robotic mode observation of the TFRM.

Figs. 3 and 4 show three examples of satellites images taken with the TFRM during the CO-VI ESA SST/SSA-PP optical observational campaign, and Fig. 5 and example of a LEO object tracking. Note that only a very small fraction of the whole TFRM 4.4ºx4.4º FoV is displayed. Also note that the telescope was tracking the satellites in all three cases.

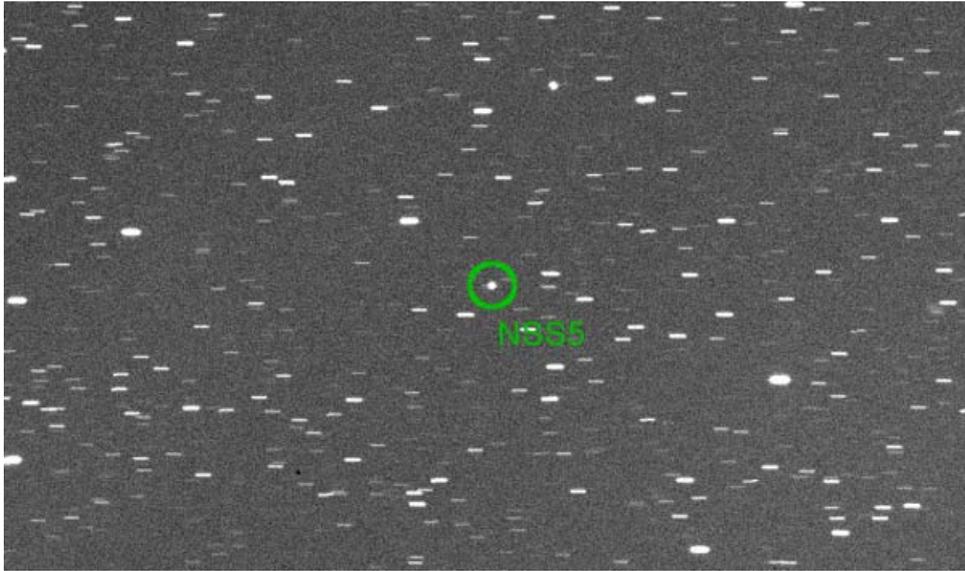

Fig. 3. A 5s exposure of the GEO satellite NSS5. Note that, as NSS5 is geostationary and the exposure time is not very long, the stars are short trailed.

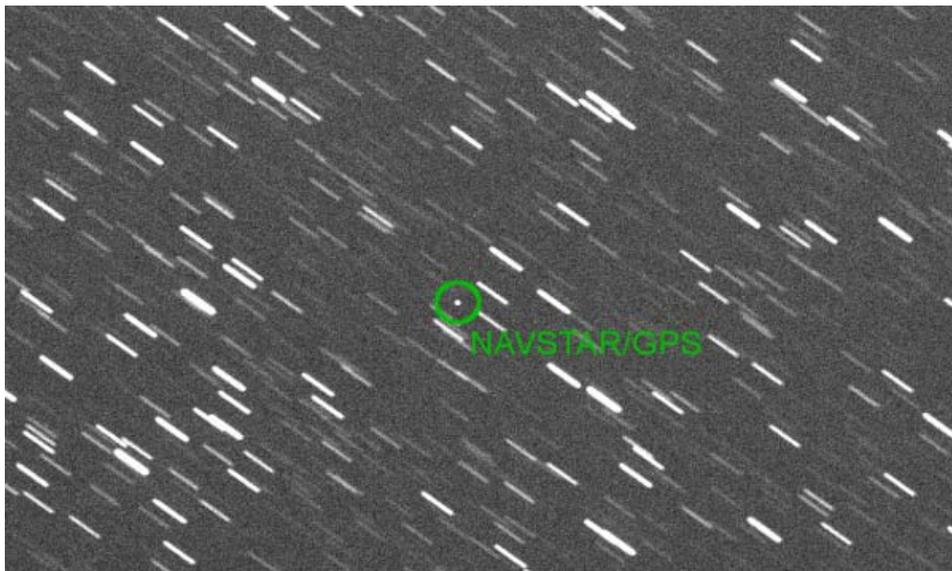

Fig. 4. A 5s exposure of the NAVSTAR 64/GPS constellation satellite. These satellites exposures were used for calibrating the GEO satellites observations. Given the closer NAVSTAR/GPS orbit than GEOs, with the same 5s exposure time, the stars are more trailed than Fig.3.

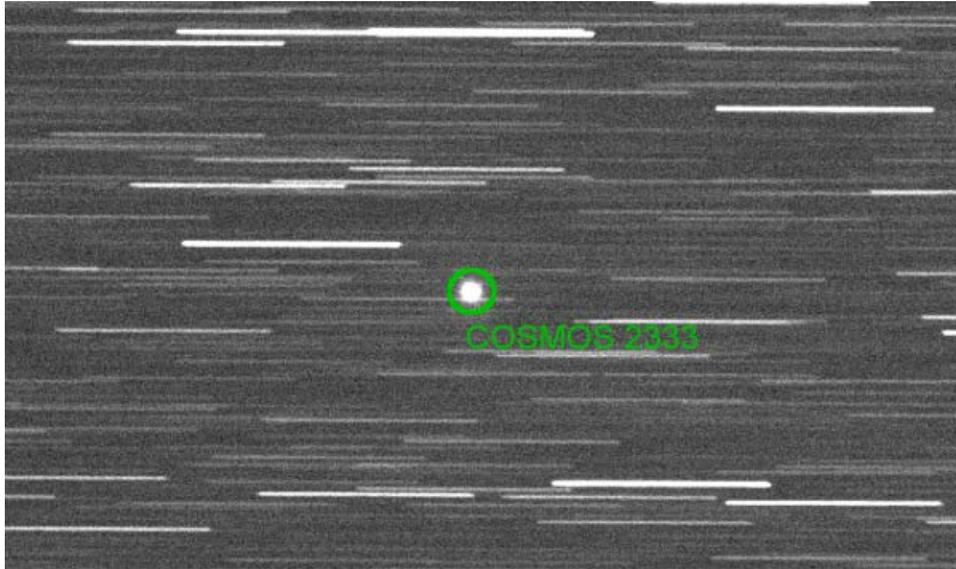

Fig. 5. A 1s exposure of COSMOS 2333. This is a LEO object, with an average orbital distance of 850 km. This explains the much longer star trails despite of the shorter exposure time with respect to Figs.3 and 4.

As an example of TFRM in GEOs objects survey performance, the TFRM has just started to participate with ISON as external collaborator [5]. Preliminary results are promising but still deserve further observations to optimize survey efficiency, accuracy and other key parameters.

Finally, it is specially rewarding and also historically noteworthy that, after its intensive refurbishment, the TFRM performance, together with the remote and robotic functionalities, allows a BNC to revisit an observational program (optical satellite tracking and space debris survey) to which was conceived.

## 5.   FIRST GEO SATELLITE TRACKING RESULTS

A figure of merit of the ESA CO-VI SST optical observations conducted at TFRM is to compute the Orbit Determination (OD) from the angular measurements. This was carried out using the Orbit Determination Tool Kit (ODTK) software package, from Analytical Graphics, Inc. (AGI)[1].

As an example, we show in Fig.6 the 2-sigma (95%) uncertainties obtained over the MSG2 satellite, with 175 angular measurements along the 4-night interval (3 Feb 2011 23:00:00 UTC to 07 Feb 2011 08:00:00 UTC) in which the satellite was not maneuvered.

---

[1] This licensed software is used by ROA in collaboration with the Instituto Nacional de Técnica Aeroespacial (INTA).

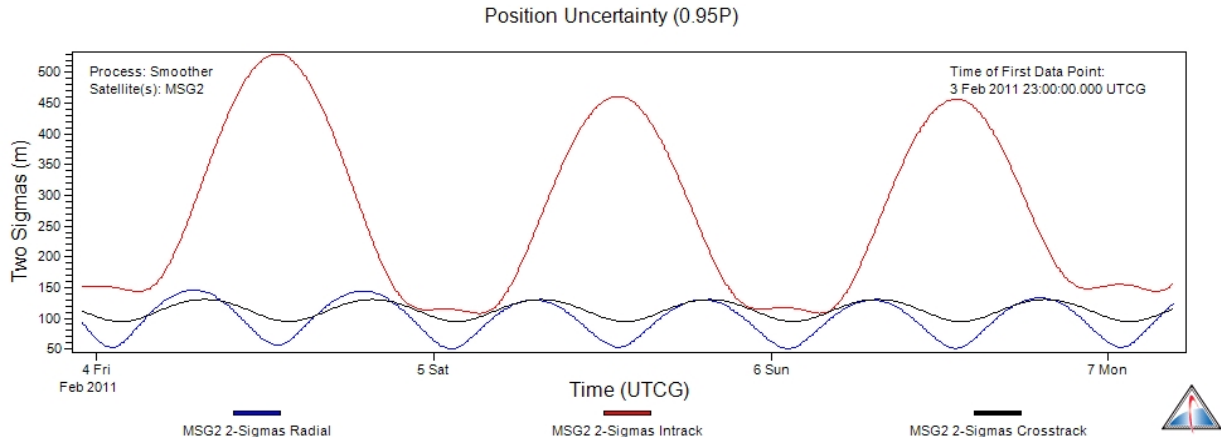

Fig. 6. 2-sigma (95%) uncertainties in the position of the MSG2 satellite and in RIC coordinates. Note the characteristic increase of the Intrack uncertainty during daytime.

The mean uncertainties in the classical elements (semiaxis, eccentricity and inclination), are of the order of 12m, $1.8 \cdot 10^{-6}$ and $1.5 \cdot 10^{-4}$ deg respectively. These preliminary results reveal the usefulness of TFRM in GEO region tracking tasks.

After some hardware and software upgrades which we are currently being carried out at the TFRM, we plan to calibrate our GEO tracking performance through the intensive observation of known GEO satellites and the comparison with the orbital elements provided by the Satellite Control Agencies [7].